\documentclass[preprint,aps]{revtex4}
\begin{document}

\title{Systematic model behavior of adsorption on flat surfaces}
\author{R.A. Trasca, M.W. Cole and R. D. Diehl}
\affiliation{Department of Physics, The Pennsylvania State University, University Park, PA 16802}

\begin{abstract}

A low density film on a flat surface is described by an expansion involving the
first four virial coefficients. The first coefficient (alone) yields the
Henry's law regime, while the next three correct for the effects of interactions.
The results permit exploration of the idea of universal adsorption behavior,
which is compared with experimental data for a number of systems.

\end{abstract}

\maketitle

\section{Introduction}

Nearly half a century ago the first adsorption experiments were carried out that
yielded behavior interpreted in terms of two-dimensional (2D) theoretical
models \cite{steele1}. The agreement with those models was a consequence of the fact that high
surface area forms of graphite present extended, uniform and flat areas, as is
assumed in the models. Subsequently, a wide variety of techniques have found similar 
behavior on other surfaces \cite{milton1}. With an abundance of such data
comes the
opportunity to address questions about universality of the behavior. If
universal relations are appropriate, it facilitates the planning and
interpretation of new experiments.

In this paper, we address this question and several others. To be specific, we
examine the nature of low coverage adsorption, using the technique of the
virial expansion. In this case, the expansion includes both gas-surface and
gas-gas interactions. Fortunately, virial coefficients incorporating the
interparticle interactions are available, facilitating this work. Hence, the
principal effort reported here involves using these coefficients along with a
simple description of the motion perpendicular to the surface. The results
include numerical values of the pressure threshold for adsorption and the
relationship between the isosteric heat of adsorption and the well depth of the
gas-surface interaction. In addition, the validity of the expansion is tested
by comparing the predictions with experimental data for diverse adsorption systems.

\section{Description of model}

The simplest model of a monolayer film in coexistence with vapor considers it as
a classical noninteracting gas in an external field, the interaction
with the substrate \cite{milton1}. Assuming that the adsorbate atoms move freely across the surface, the 
potential energy of adsorption is a function of the normal coordinate, z, alone. An 
accurate representation of many adsorption potentials, sufficient to describe the 
energetics and dynamics in the vicinity of the minimum, is:
\begin{equation}
V(z)=-D+\kappa z^{2}/2
\end{equation}
where $\kappa$ is the force constant, $D$ is the well depth and the equilibrium position is $z=0$. The 
solution for the number density of a noninteracting gas in the potential $V(z)$ 
follows as:
\begin{equation}
n(z)=n(\infty) \exp[-\beta V(z)]
\end{equation}
where $\beta=1/(k_B T)$, T is the temperature and $n(\infty)$ is the vapor density as $z \rightarrow \infty$, 
where the true potential energy (but not (1)) vanishes. 
Then, the film coverage on a surface of area A is given by the expression \cite{footnote1}:
\begin{equation}
N=A \int n(z) dz= n(\infty) \int \exp[-\beta V(z)] dz
\end{equation}
With the assumed quadratic dependence of $V$ on $z$ and the equation for an 
ideal gas $n(\infty)=\beta P$, the film's two-dimensional density $\theta=N/A$ is given by:
\begin{equation}
\theta= \beta P \sqrt(\frac{2\pi}{\beta \kappa}) \exp(\beta D)
\end{equation}
This Arrhenius form is the generic behavior of adsorption isotherms,
with an activation energy equal to the well depth. Thus, thermodynamic techniques
can be employed for finding the pressure-temperature conditions necessary for adsorption
and probe the well depth \cite{webb1,webb2}. However, the noninteracting gas model
is applicable only at very low densities.

An improved model should take into account interactions between the adsorbate atoms, the effect of which
usually predominates over effects of quantum statistics (except at very low temperatures
for light and weakly interacting gases) \cite {guo,siddon,Pathria,steele, maryjo}.
This improvement is provided by the virial expansion, a series expansion whose
leading term denotes the ideal-system results, while subsequent terms
provide corrections arising from the interparticle interactions.
For an isoenergetic, or structureless, substrate for which the motion is
well confined to the plane $z=0$, the problem becomes essentially 2D. 
The virial expansion of the equation of state
of a two-dimensional gas can be written in the form \cite{morris}
\begin{equation}
\beta \Pi = \theta + B_{2D} \theta ^2 + C_{2D} \theta ^3 + D_{2D} \theta^4 ...
\end{equation}
where $\Pi$ is the 2D (spreading) pressure of the monolayer film. 
$B_{2D}$,$C_{2D}$ and $D_{2D}$ are
the 2D second, third and fourth virial coefficients, respectively. In the 2D case,
assumed here, the second virial
coefficient is related to the
two-body interaction through the equation
\begin{equation}
B_{2D}= -\pi \int_0^{\infty} dr {r [\exp(-\beta u(r))-1]}
\end{equation}
where $r$ is the intermolecular separation and $u(r)$ is the pair potential. The van der Waals (VDW) theory
is sometimes used to find an approximation of the second virial coefficient
\begin{equation}
 B_{2D}^{VdW} \approx 1/2 \int_0^\sigma d^2 r + \beta /2 \int_\sigma ^\infty u(r) d^2 r 
\end{equation}
In the case of a Lennard-Jones type of interaction,
\begin{equation}
u(r)=4\epsilon [(\frac{\sigma}{r})^{12}-(\frac{\sigma}{r})^{6}]
\end{equation}
where $\epsilon$ and $\sigma$
are the interparticle well depth and hard-core diameter, respectively. Using this
interaction, the second virial coefficient in the VDW approximation is
\begin{equation}
B_{2D}^{VdW}=\frac{\pi \sigma^2}{2}-\beta \frac{3\pi \epsilon \sigma^2}{5}
\end{equation} 
We computed both the "real" $B_{2D}$ and its VDW approximation, as indicated in $Fig. 1$.
Note that they differ
appreciably at any T. However, one can shift the VDW approximation by adjusting
the parameters $a$ and $b$ in the equation $B_{2D}=b-\beta a$. The adjustment to $a$
and $b$ was done by modifying the
interparticle well depth to $\epsilon_{VDW}=1.3 \epsilon$, a correction which 
improves the accuracy of the VDW approximation as seen in the figure. In this way, we got a good 
fit to $B_{2D}$ at intermediate T, corresponding to the range of many experiments. At very high T,
however, both VDW curves converge to a finite value, while the real $B_{2D}$ goes to zero.

The third virial coefficient is related to the
two-body interactions in clusters of three particles and is defined by the equation
(in the 2D approximation):
\begin{equation}
C_{2D}= -\frac{1}{3} \int_0^{\infty} dr_{2} dr_{3} [\exp(-\beta u(r_{12}))-1][\exp(-\beta u(r_{13}))-1][\exp(-\beta u(r_{23}))-1]
\end{equation}
This expression omits any explicit three-body interactions \cite{gray}.
Details of the computation and tables of numerical values of $B_{2D}$ and 
$C_{2D}$ can be found in the original sources \cite{morris,steele}. 

The equation of state relates the spreading pressure $\Pi$ to $\theta$ and $T$. However,
in an experiment one usually manipulates $P$ and $T$. Therefore,
we seek a relationship $P(\theta,T)$.
This can be accomplished by employing the equilibrium condition between the gas phase and
the adsorbed film
\begin{equation}
\mu_{gas}=\mu_{film}
\end{equation}
where $\mu$ is the chemical potential in the respective phase.
The chemical potential
of the gas (assumed spinless) phase is well known to be 
\begin{equation}
\beta \mu_{gas}=ln(\beta P \lambda^3)
\end{equation}
where
\begin{equation}
\lambda=\frac{h}{(2\pi m kT)^{1/2}}
\end{equation}
is the thermal wavelength
of the particles. A derivation of $\mu_{film}$ appears in the appendix, together with a
discussion of the limitations
of this approach: 
\begin{equation}
\beta\mu_{film}=\ln(\theta \lambda^2)-\beta D+2 B_{2D}\theta + \frac {3}{2} C_{2D} \theta^2 + \frac{4}{3} D_{2D} \theta^3
\end{equation}
Using the equilibrium
condition ($Eq. 11$), the relationship between $\theta$, $P$ and $T$ is found
\begin{equation}
\theta=\beta P \lambda \exp(\beta D)\exp(-2 B_{2D}\theta-\frac{3}{2}C_{2D} \theta^2-\frac{4}{3} D_{2D} \theta^3)
\end{equation}

A kind of universal version of this equation appears if the thermodynamic
quantities are replaced with reduced (dimensionless) quantities. Thus
$P$ can be written as a function of a dimensionless pressure ($P^*$):
$P=P^* \epsilon/\sigma^3$.
In the same way, $D=D^* \epsilon$, $k_B T=T^* \epsilon$ and $\theta=\theta^* /\sigma^2$. 
The virial coefficients can also be reduced as $B_{2D}^*=B_{2D}/\sigma^2$, $C_{2D}^*=C_{2D}/\sigma^4$
$D_{2D}^*=D_{2D}/\sigma^6$. A comparison of $Eq. 15$ and $Eq.4$ puts $\lambda$ in correspondence
with $\sqrt{2\pi/\beta \kappa}$, which will be used
instead of $\lambda$.
The force constant ($\kappa$) is taken to be $D/\sigma^2$, resulting in a universal
adsorption equation
\begin{equation}
\theta^*= P^* \sqrt{\frac{2 \pi}{T^* D^*}} \exp(D^*/T^*) \exp(-2B_{2D}^* \theta^*-\frac{3}{2} C_{2D}^* {\theta^*}^2-\frac{4}{3} D_{2D}^* {\theta^*}^3)
\end {equation}

An important question (particularly useful for planning experiments)
we address is this: what value of $P^*$ produces "significant"
adsorption? By "significant", we mean the opposite of non-negligible, arbitrarily
defined as reduced density $\theta^* \ge 0.1$. Below this density,
the virial expansion should suffice over much of the relevant range of T, so we employ it to determine this pressure threshold.
$Fig.2$ shows the resulting adsorption threshold curves for various numbers of virial terms and
for three different values of $D^*$.
Note that (for
a specific well depth) the four curves converge at high T,
since then the effect of gas-gas interactions disappears. At
intermediate T, in contrast, the effect of interactions is to enhance the adsorption, which begins at a lower
reduced pressure. The limitations of the virial expansion approach is seen in the difference
between the black and red curves. At low T, the curves diverge and the $C_{2D}$
correction is large, as is that of $D_{2D}$ (for $T^* < 0.45$). Thus,
we conclude that higher order corrections cannot improve the convergence of the virial series,
since the virial coefficients are themselves divergent for $T^* < 0.45$ \cite {morris, steele, glandt}.
This divergence is consistent with the 2D critical temperature of different gases ($T_c^* \simeq 0.5$)
\cite {ref}.

The curves presented in $Fig.2$ can be applied to treat specific adsorbates
and substrates. The standard LJ parameters are well known in the literature \cite{watts,maryjo}.
However, when comparing our results with experiments, we assessed effects of the reduction of
LJ well depth due to substrate screening effects (the McLachlan interaction) 
\cite {screen,mc}. Table 1 presents both sets of parameters.

Adsorption experiments allow one to find the isosteric heat, which is
a rough measure of the gas-substrate well depth. The relationship between the
isosteric heat and well depth is not straightforward. 
The isosteric heat is
\begin{equation}
Q_{st}=-(\frac{\partial \ln P}{\partial \beta})_N
\end{equation}
The adsorption equation ($Eq. 15$) allows us to find the P-T relation,
from which one obtains
\begin{equation}
\beta Q_{st}=\frac{1}{2}+\beta D - 2 \theta \frac{d B_{2D}}{d \ln \beta} - \frac{3\theta^2}{2}\frac{d C_{2D}}{d \ln \beta} 
\end{equation}
In dimensionless form
\begin{equation}
Q_{st}^* - D^*=\frac{1}{2} T^* + 2 (T^*)^2 \frac{d B_{2D}^*}{d T^*} \theta^*+ \frac{3}{2} (T^*)^2 \frac{d C_{2D}^*}{d T^*} (\theta^*)^2
\end{equation}
where $Q_{st}^*=Q_{st}/\epsilon$ is the reduced isosteric heat. $Fig. 3$ shows the 
difference between the reduced isosteric heat and the well depth, as a function
of T in three cases:
noninteracting gas, interacting gas including only $B_{2D}$
and interacting gas including both $B_{2D}$ and $C_{2D}$. For specificity, we consider
the case of $\theta^* =0.1$. Let us
focus first at low T: the difference between $Q_{st}^*$ and $D^*$ is small for
the noninteracting model ($T^*$), but for the interacting gas the isosteric heat and well depth
differ more due to the (overall attractive) gas-gas interactions. (Again,
caution must be applied with the derivatives of virial coefficients, divergent for $T \le 0.5$.)
For intermediate values of $T^*$,
this difference is of the order
of the well depth $\epsilon$, which is particularly important in cases when
$\epsilon \ge D$. At $T^* \approx 2$, the interacting and
noninteracting models yield quite different values of $Q_{st}$. This can be seen easily for
the second coefficient, computed in the VDW approximation,
\begin{equation}
(T^*)^2 \frac{d B_{VDW}^*}{d T^*}=\frac{3\pi}{5}
\end{equation}
When $T \rightarrow \infty$, however, $B_{VDW}$ is still
finite while $B_{2D} \rightarrow 0$. Thus, the noninteracting and interacting models
converge at high T to the limit $Q_{st}^*-D^*=T^*/2$.

\section{Comparison with experiments}

A relevant assessment of these expressions is possible for adsorption on graphite and 
Ag (111) substrates,
since there is a large body of experimental data for these surfaces. 
The data were obtained from electron diffraction
studies of the adsorption of various gases.
During these experiments, gas was adsorbed
under quasi-equilibrium conditions and the intensity of the diffraction spots
from the substrate was measured.  The attenuation of these diffraction
beams provides a measure of the amount of gas adsorbed; in all of these
cases, distinct "steps" were observed in the resulting equilibrium isobars 
or isotherms, corresponding to adsorption of the first layer.  (Subsequent steps
corresponding to subsequent layers were also observed in most cases, but
those are not relevant to this paper.)  By studying this behavios over
a range of pressure or temperature, it is possible to produce an
"isostere" curve in the P-T plane
for any chosen coverage.  The coverages chosen in the 
experimental studies cited here were the half-monolayer and monolayer coverages.
Because of the limitation posed by the mean-free path of electrons, gas 
pressures higher than about $10^{-3}$ mbar were inaccessible in these experiments,
and therefore the range of data available for comparison is the pressure
range between about $10^{-11}$ mbar and $10^{-3}$ mbar.

As sources
for our well depth values, we used Refs. \cite{vidali} for graphite and 
\cite{bruch} for Ag(111). 
Results of our calculations are shown (using the screened LJ parameters from Table 1) in 
$Fig.4$ (for graphite) and $Fig. 5$ (for Ag(111)).
The experimental data were taken from various references 
\cite{N2/gr, Ar/gr, Kr/gr, H2/gr, Ne1/gr, Ne2/gr,Xe/gr,Ar_Kr/Ag, Xe/Ag, N2/Ag}. 
The calculated curves are in overall good agreement with the experimental points, taking into
account the approximations we made to achieve a simple, universal curve. However, we noticed 
that the calculations are sensitive to the gas interaction parameters; the standard LJ parameters yield
less agreement than those of the screened interaction. The discrepancies
between the calculated curves and some data points may be due to the fact that our calculations 
are performed at coverages much lower than a monolayer, while experiments usually 
tabulate the pressure and temperature
values related to the completion of the monolayer. However, experimentally, the low coverage
pressure (at the onset of adsorption) does not usually differ appreciably from the monolayer
completion pressure. As mentioned above, the virial 
expansion is not convergent for $T< 0.4 \epsilon$; therefore
for gases with large $\epsilon$, some experimental points
may be in this range and thus other theoretical methods (Monte Carlo simulations)
should be applied.

\section{Conclusions}

Virial expansion calculations were performed in order to find the pressure-temperature
threshold of adsorption on various surfaces assumed to be atomically smooth and perfect
so they conform to the 2D approximation. 
Employing the equilibrium condition between the film and vapor, an equation
relating $P$ to $\theta$ and $T$ was found and 
reduced to a universal equation. A comparison between the noninteracting
and interacting models showed that
the effect of interactions is to enhance the onset of adsorption, which occurs at lower pressure
than in the noninteracting case. At high T, there is little difference between the
noninteracting and interacting curves, while at intermediate T, the difference between the
them is larger. At very low T, the virial coefficients are divergent. 
The computed threshold of adsorption was tested for various gases
adsorbed on graphite and Ag(111). The calculated curves are in good agreement with the
experimental. However, some of the experimental points lie in a
range where the virial expansion diverges, so the comparison is incomplete.
Finally,
we addressed the relationship between the isosteric heat and the well depth.
Due to interactions, these quantities differ appreciably at low T, while at high T the
noninteracting and interacting cases converge to the same values.

We thank Mary Jo Bojan and Bill Steele for helpful information and the National 
Science Foundation for support of this research.

\section{Appendix}

 The film's chemical 
potential can be found from the thermodynamic equation 
\begin{equation}
\beta \mu_{film}=-\left (\frac {\delta \ln Q_N}{\delta N}\right)_T
\end{equation}
where $Q_N$ is the N-particle partition function of the film. The partition function
for a noninteracting 2D gas can be factorized
into $Q_{2D}$, associated with
motion parallel to the surface, and $Q_z$, associated with motion
perpendicular to the surface. Then, the partition function can be written as 
\begin{equation}
Q_N= \frac{1}{N!} (\frac{A}{\lambda^2}Q_{z})^N
\end{equation}
where A is the area of the film. The molecules'
motion perpendicular to the surface depends on the gas-surface interaction potential,
here taken as $V(z)=-D+kz^2/2$.
Then, the quantum partition function for the $z$ motion is
\begin{eqnarray}
Q_{z}^{quan}&=& \sum_{n=0}^\infty \exp[-\beta (-D+\hbar \omega (n+1/2))]
\nonumber\\
&=& \frac{\exp(\beta (D-\hbar \omega/2))}{1-\exp(-\beta \hbar \omega)}
\nonumber\\
&=& \frac{\exp(\beta E_b)}{1-\exp(-\beta \hbar \omega)} 
\end{eqnarray}
Here, $\hbar \omega /2$ is the zero point energy,
$ 1-\exp(-\beta \hbar \omega)$ comes from excitations to higher levels
and $E_b=D-\hbar \omega /2$ is the binding energy. 
Classically, instead
\begin{eqnarray}
Q_z^{cl} &=&\int \frac {dz}{\lambda} \exp[-\beta (-D+kz^2/2)]
\nonumber\\
&=& \frac {\exp (\beta D)}{\beta \hbar \omega}
\end {eqnarray}
Thus,
the ratio of $Q_z^{cl}$ to $Q_z^{quan}$
\begin{equation}
R=\frac{Q_z^{cl}}{Q_z^{quan}}=\frac{2\sinh(\beta \hbar\omega/2)}{(\beta \hbar \omega)}
\end{equation}
A high T expansion yields $R\approx 1+ (\beta \hbar \omega)^2/24$. This expansion works
well up to $\beta \hbar \omega =1$, at which point $R=1.04$.
At $\beta \hbar \omega =1.5$, $R=1.1$. Since one takes the
logarithm to get the chemical potential, a $10 \%$ difference in $Q_z$ is not 
significant. For the high T regime, $\beta \hbar \omega < 1.5$.
Therefore the classical approximation can be used for most purposes, and it yields:
\begin{equation}
\beta \mu_{film} = \ln(\theta \lambda^2)-\beta D+ \ln(\beta \hbar \omega)
\end{equation}
For $\beta \hbar \omega$ of order 1, the last term in the above equation is negligible,
and it makes only logarithmic corrections elsewhere. 
Thus, the film chemical potential at low density is essentially 
\begin{equation}
\beta \mu_{film} \simeq \ln(\theta_0 \lambda^2)-\beta D
\end{equation}
where $\theta_0=N/A $ is the film coverage at low densities. In deducing this equation
the film was supposed to be a 2D noninteracting gas. 
The calculation of the chemical potential, including interactions
through the fourth virial coefficient, can be found with the 2D 
Gibbs-Duhem equation
\begin{equation}
\left (\frac{\partial \mu}{\partial \Pi}\right)_T=\frac{1}{\theta}
\end{equation}
Integrating
\begin{equation}
\left (\frac{\partial \mu}{\partial \theta}\right)_T =\left(\frac{\partial \mu}{\partial \Pi}\right)_T \left(\frac{\partial \Pi}{\partial \theta}\right)_T \simeq (1+2B_{2D}\theta +3C_{2D}\theta^2+4D_{2D}\theta^3)/(\theta\beta)
\end{equation}
from a very low density $\theta_0$ (where the gas can be considered 
noninteracting) to a higher density ($\theta$), the film chemical potential becomes
\begin{equation}
\beta\mu_{film}=\ln(\theta \lambda^2)-\beta D+2 B_{2D}\theta + \frac {3}{2} C_{2D} \theta^2 +  \frac {4}{3} D_{2D} \theta^3
\end{equation}

\newpage

\begin{center}
\begin{table}[h]
\caption{Standard and screened LJ parameters. Notice the well depth
($\epsilon$) reduction due to screening \cite{screen,mc} when adsorbed on graphite
and Ag (111). The change of $\sigma$ due to screening is not significant, thus
is not shown.}

\setlength{\tabcolsep}{0.4cm} \renewcommand{\arraystretch}{1.5}
\begin{tabular*}{12 cm}[t]{|c|c|c|c|c|}\hline

$gas$& $\epsilon_{standard}$ (K)& $\epsilon_{screened}^{Gr}$ (K)&
$\epsilon_{screened}^{Ag}$ (K)& $\sigma_{st}(\mathring{A})$  \\ \hline

$Xe$&   221&    214&   185&   4.10 \\

$Kr$&   171&    150&   135&   3.60  \\

$Ar$&   120&    110&   96.7&  3.40  \\

$H_2$&   37.0&    27.3&   -&   3.05  \\

$Ne$&    35.6&    33.8&  -&   2.75 \\

$N_2$&   36.4&    -&    -&   3.32 \\

$CO$&   49.6&    -&    -&    3.14  \\

$CH_4$&   148&    -&   -&   3.45 \\

$He$&   10.2&    -&  -&   2.56   \\ \hline

\end{tabular*}
\end{table}
\end{center}

\newpage
\section {Figure captions}

1. The reduced second virial coefficient as a function of reduced temperature. The
triangles correspond to an exact calculation (Eq.6), the circles use the VDW 
approximation (Eq. 9) and the square curve is obtained by shifting the VDW curve
to the right (corresponding to an increase in gas well depth: $\epsilon _{VDW}=1.3 \epsilon$).
Inset shows low T behavior.

2. Reduced pressure at threshold for adsorption as a function of reduced temperature 
for three reduced well-depth values $D^*$. 
The squares correspond to a non-interacting gas; the circles take into account
interactions through $B_{2D}$, the triangles include both $B_{2D}$ and $C_{2D}$ and the
plus curves include the fourth virial term.
In each case, little or no adsorption occurs below the curve.

3. Difference between the reduced isosteric heat and well depth as a function of reduced
temperature in three cases at $\theta^*=0.1$: noninteracting gas (triangle), interacting gas through $B_{2D}$
(square) and interacting gas with $B_{2D}$ and $C_{2D}$ (circle).

4. Threshold pressure of various gases on graphite compared with various experiments 
discussed in the text. The dashed curves take into account $B_{2D}$ and the full curves
include both $B_{2D}$ and $C_{2D}$.

5. Threshold pressure of various gases on Ag(111) compared with various experiments 
discussed in the text. The dashed curves take into account $B_{2D}$ and the full curves
include both $B_{2D}$ and $C_{2D}$.


\begin{references}
\bibitem{footnote1}
For the present paper, we do not distinguish between the film coverage and the thermodynamic
excess coverage, obtained by substracting from $Eq. 3$ the integral in the case $V=0$; the
latter requires a specification of the film domain, which is ambiguous.
\end{references}
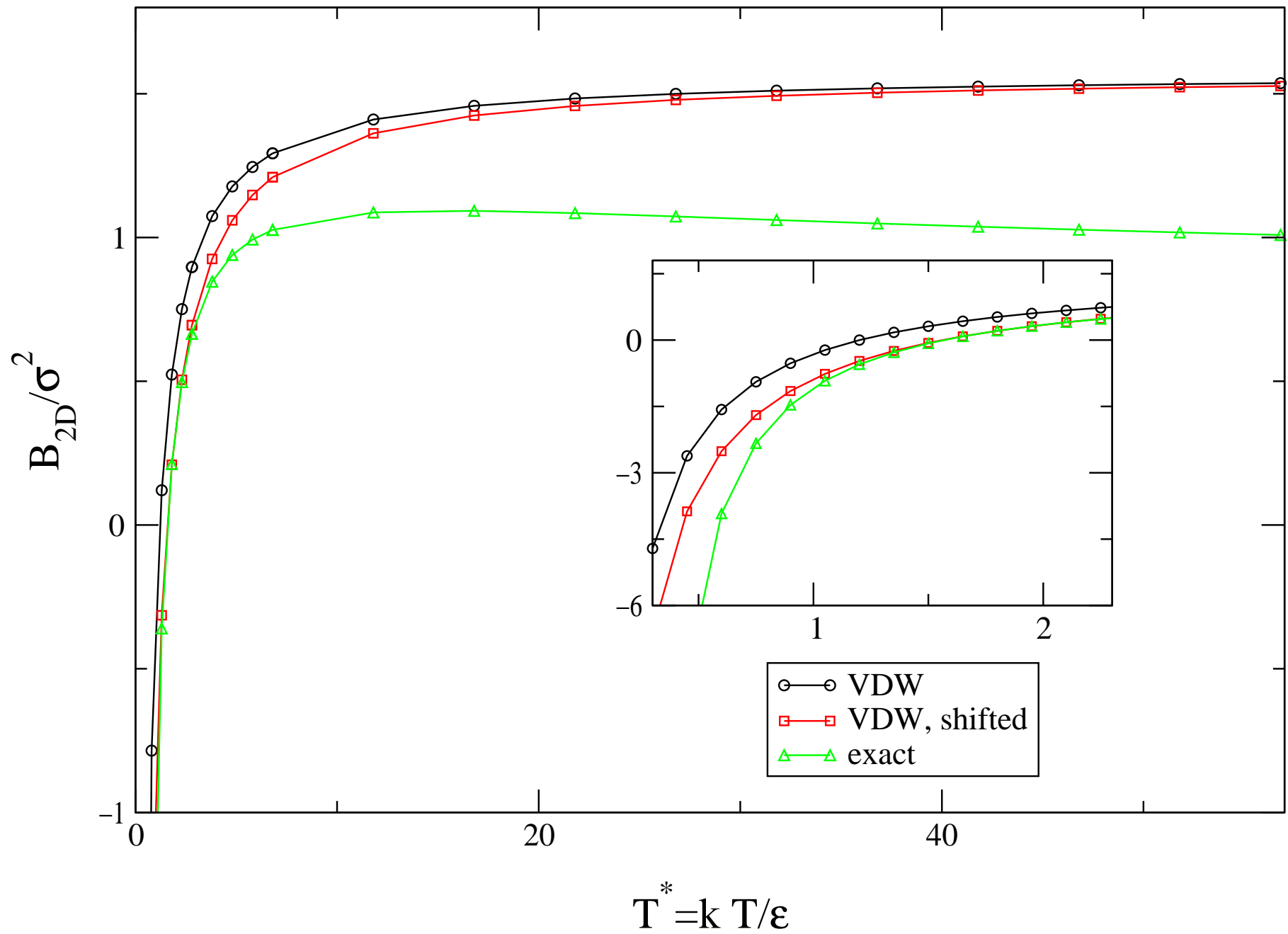
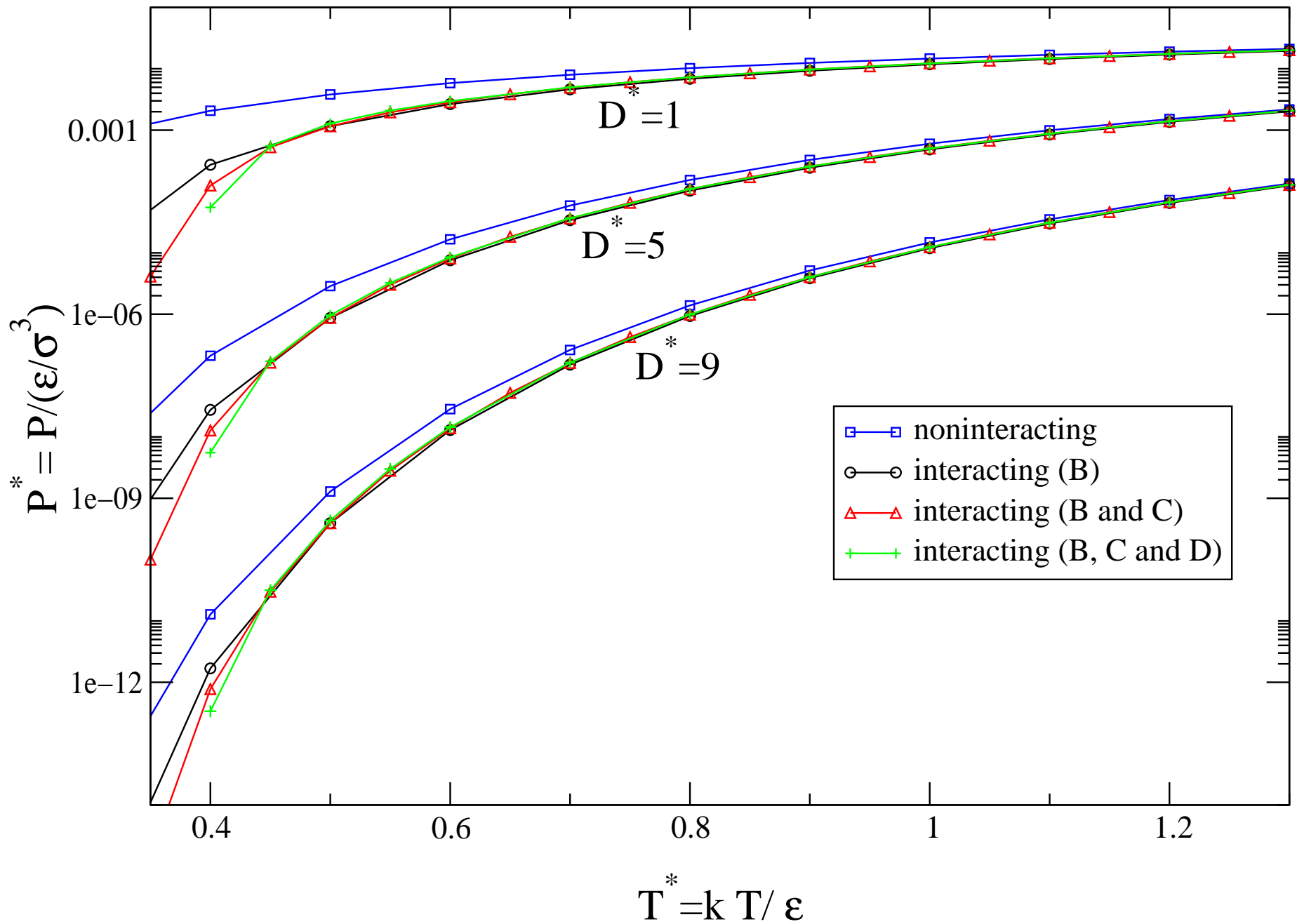
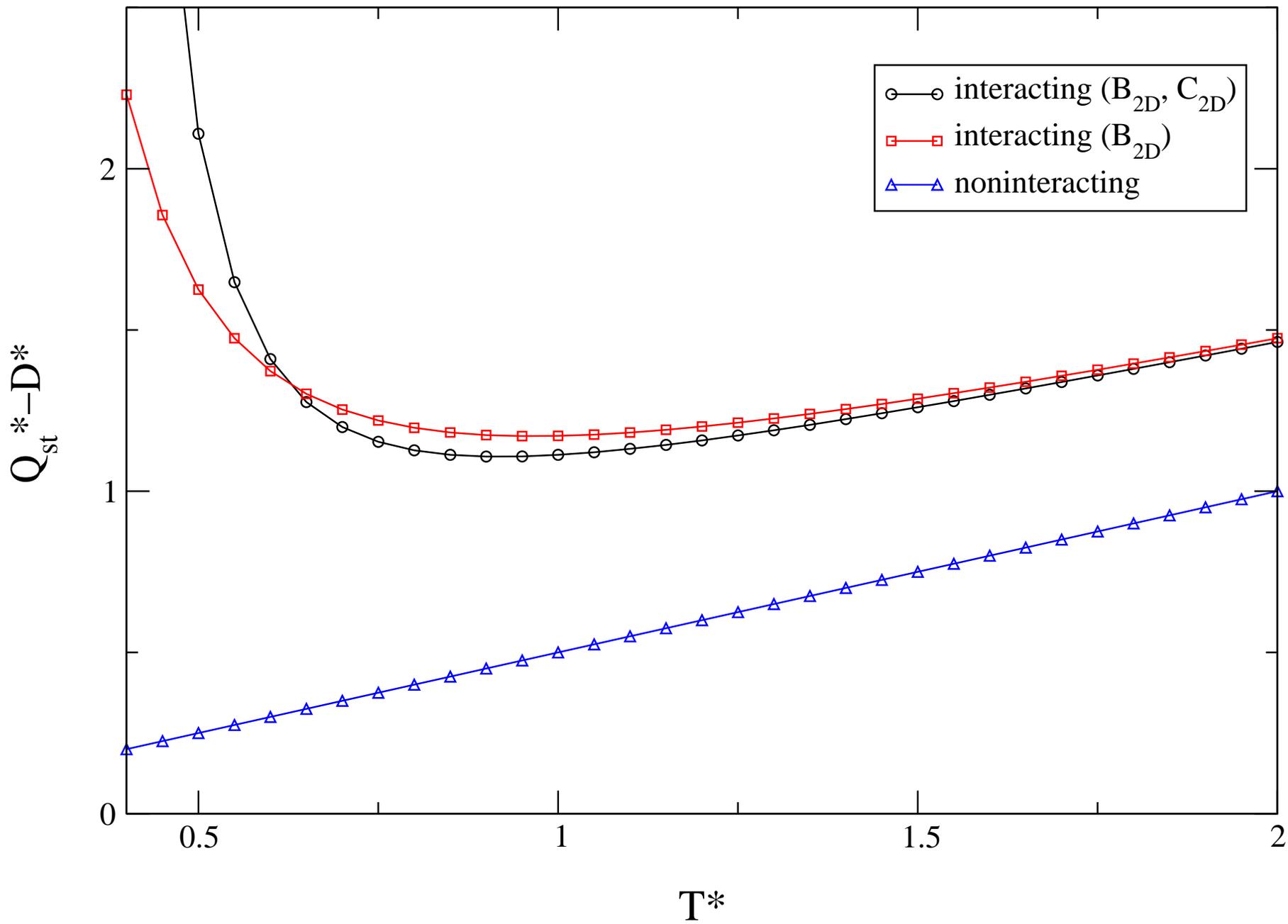
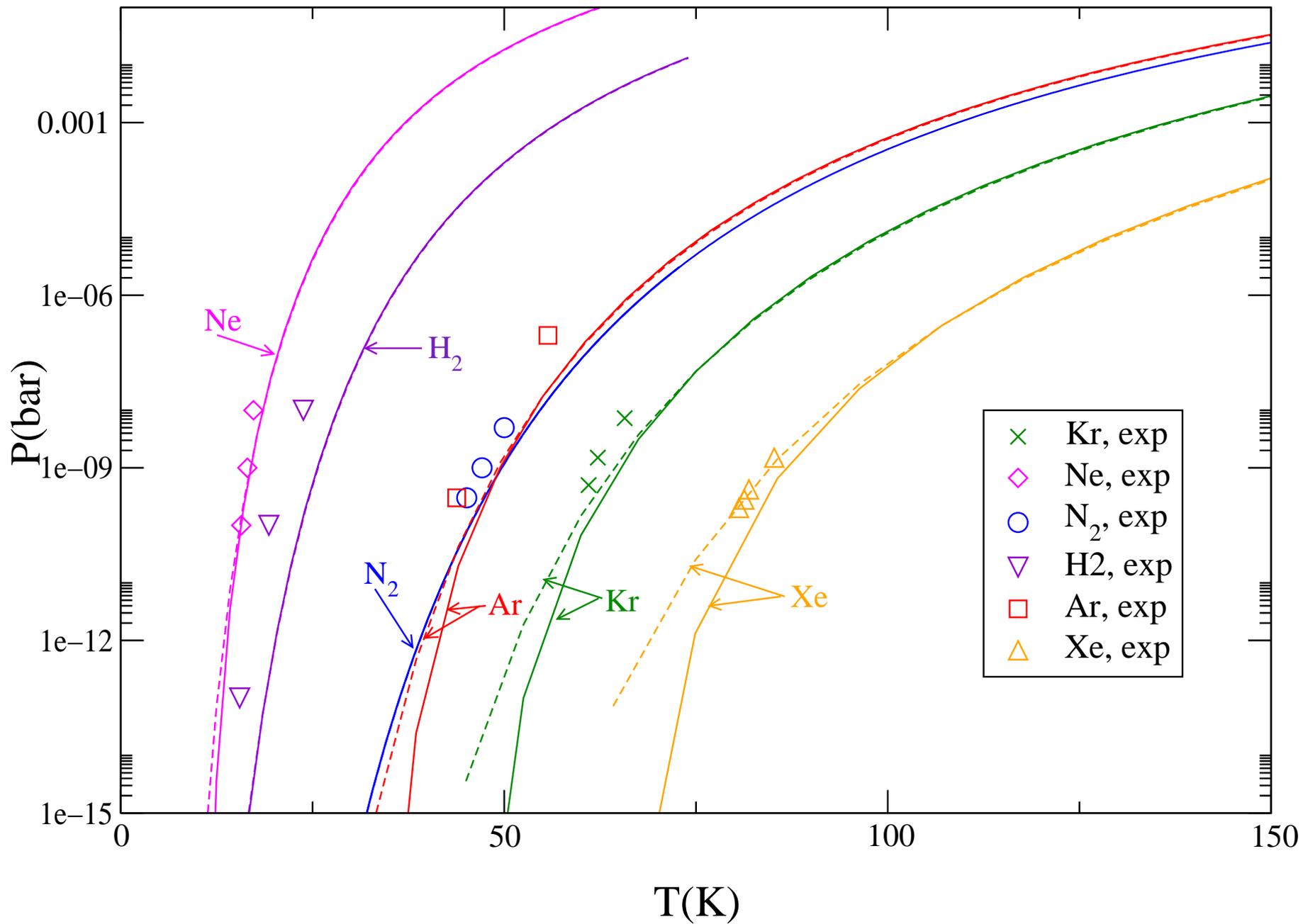
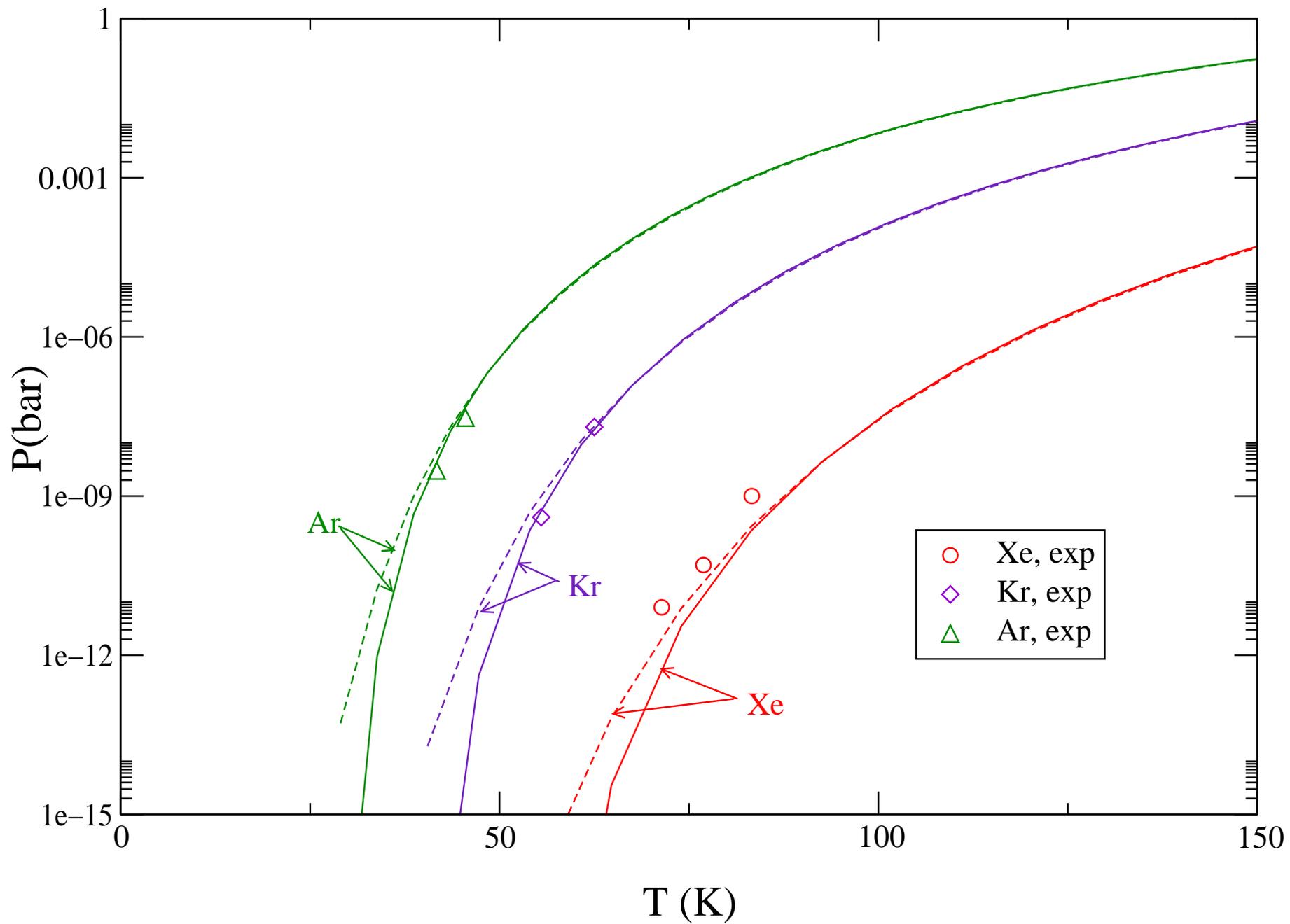
\end{document}